\documentclass[aps,pra,twocolumn,superscriptaddress]{revtex4-2}
\usepackage{amsmath,amsfonts,amssymb}
\usepackage{graphicx}
\usepackage{epstopdf}
\usepackage{epsfig}
\usepackage[normalem]{ulem}
\usepackage{multirow}
\usepackage{makecell}
\usepackage{array}
\usepackage{booktabs}
\usepackage{mathtools}
\usepackage{bm}
\usepackage{color}
\usepackage{soul,xcolor}
\usepackage{array}
\usepackage{booktabs}
\usepackage{braket}
\usepackage{mathrsfs}
\usepackage[utf8]{inputenc}

\usepackage{tabularx}
\newcolumntype{Y}{>{\centering\arraybackslash}X}

\usepackage{blindtext}
\usepackage{comment}
\definecolor{lightblue}{rgb}{0.3, 0.3, 0.90}
\definecolor{darkblue}{rgb}{0.0, 0.2, 0.5}
\definecolor{lightred}{rgb}{1.0, 0.6, 0.6}
\definecolor{darkred}{rgb}{0.8, 0.0, 0.0}

\begin{document}

\title{Deficiency of equation-finding approach to data-driven modeling of dynamical systems}

\author{Zheng-Meng Zhai}
\affiliation{School of Electrical, Computer and Energy Engineering, Arizona State University, Tempe, Arizona 85287, USA}

\author{Valerio Lucarini}
\affiliation{School of Computing and Mathematical Sciences, University of Leicester, Leicester LE1 7RH, UK}

\author{Ying-Cheng Lai} \email{Ying-Cheng.Lai@asu.edu}
\affiliation{School of Electrical, Computer and Energy Engineering, Arizona State University, Tempe, Arizona 85287, USA}
\affiliation{Department of Physics, Arizona State University, Tempe, Arizona 85287, USA}

\date{\today}

\begin{abstract}

Finding the governing equations from data by sparse optimization has become a popular approach to deterministic modeling of dynamical systems. Considering the physical situations where the data can be imperfect due to disturbances and measurement errors, we show that for many chaotic systems, widely used sparse-optimization methods for discovering governing equations produce models that depend sensitively on the measurement procedure, yet all such models generate virtually identical chaotic attractors, leading to a striking limitation that challenges the conventional notion of equation-based modeling in complex dynamical systems. Calculating the Koopman spectra, we find that the different sets of equations agree in their large eigenvalues and the differences begin to appear when the eigenvalues are smaller than an equation-dependent threshold. The results suggest that finding the governing equations of the system and attempting to interpret them physically may lead to misleading conclusions. It would be more useful to work directly with the available data using, e.g., machine-learning methods.

%Tests of simple physical systems indicate that significant physical insights are needed to correctly interpret the equations. The finding suggests that extreme caution be exercised to interpret the equations produced by SINDy. 

\end{abstract}
\maketitle

In physical science, a methodology of biblical importance is developing a 
quantitative model by extracting a set of governing equations from experimental 
data. This is particularly relevant to chaotic 
systems~\cite{GHYS:1990,DGSY:1994,CCFV:2012} whose hallmark is sensitive dependence
on initial conditions~\cite{Ott:book}.
An essential criterion is that such equations should not depend on the 
specific measurements. For example, if an experiment is performed on a system of 
interest at two different times at which the environment and the external 
disturbances on the system could be different, it is generally expected and often 
tacitly assumed that the resulting ``inferred'' quantitative model be the same. 
The observation independence of the extracted system equations is a fundamental 
principle in science.

We report a violation of this principle for complex nonlinear dynamical systems
when attempting to find the governing equations from data. In particular, we find 
that many well studied chaotic systems, such as the Lorenz system~\cite{Lorenz:1963}, 
exhibit the surprising behavior that the governing equations found from 
data depend on the procedure used for measuring measurement of the time series. 
Consider the situation where two sets of measurement are performed under slightly 
different conditions, resulting in time series that are in principle statistically 
equivalent but in fact differ in their details. For example, due to external 
disturbance or instrumental glitches from time to time, the measured time series 
could contain missing data points, so two equally long independent measurements 
of the same system at two different times could result in time series with a 
different set of missing data points. Taking advantage of machine learning to 
reconstruct the time series through missing-data imputation and then finding the 
governing equations using some much-used sparse-optimization 
method~\cite{WYLKG:2011,Lai:2021} such as SINDy (Sparse Identification of 
Nonlinear Dynamics)~\cite{BPK:2016}, we find that the discovered equations can 
have drastically different forms. Indeed, the discovered equation can feature 
entirely new terms compared to the ``correct'' evolution equations, exclude 
some of the existing terms, or have terms with very different coefficient or 
even opposite sign. Given a chaotic system, in principle an 
infinite number of governing-equation sets of different mathematical forms can be 
found. The striking phenomenon is that all these different forms of equations 
produce the remarkably similar chaotic attractors in terms of, e.g., the Lyapunov 
exponents~\cite{Ott:book,BCFV:2002}, fractal dimension, and Kullback–Leibler (KL) 
divergence~\cite{Pardo:book}.

To understand the origin of the differences among the distinct sets of inferred 
equations from the same system and to characterize them, we resort to Koopman 
analysis~\cite{Koopman:1931}. In the Koopman theory, the basic idea is that a 
nonlinear system can be transformed into an infinite-dimensional linear dynamical 
system by shifting the perspective from looking at the evolution of the orbit 
of the system to studying the evolution of 
observables~\cite{Mezic2005,Budisic2012,brunton2022data}. The goal is to find 
eigenvalues and eigenvectors of the Koopman operator and use such a decomposition 
to explain the properties of the system, predict its behavior and, as shown 
recently, predict its response to perturbations~\cite{Zagli2024,Lucarini2025}. 
Following extensive research activities in functional analysis, algorithmics, 
and dynamical systems theory, efficient and accurate methods are available to 
perform data-driven Koopman analysis of a wide variety of systems, with recent 
progress ensuring the applicability of the strategy also beyond low-dimensional 
cases~\cite{Colbrook2024Multi}. 
%While it is generally extremely difficult to find the explicit form of the Koopman operator, in low dimensions the transform can be done numerically based on data generated by the system equations. The eigenvalue  spectrum associated with the Koopman operator can then be calculated. 
We find here that the Koopman spectra from the seemingly very different inferred 
evolution equations agree with each other for a large number - several dozens - 
of dominant eigenvalues. Differences emerge when considering eigenvalues that 
are smaller than a threshold that depends on the equation details. 

Intuitively, if two distinct evolution equations generate nearly identical chaotic 
attractors and Koopman representations, the associated velocity fields restricted 
to the attractor should exhibit close agreement. Our analysis confirms this 
expectation: the velocity fields coincide to high accuracy across most of the 
attractor, with discrepancies arising only on small subsets where sudden, spike-like 
deviations occur. Owing to their local and impulsive character, these deviations 
exert negligible influence on long-term statistical quantities such as Lyapunov 
exponents, as well as on the dominant structures of the Koopman representation. 
Agreement on the leading Koopman modes further indicates that correlations between 
generic observables and responses to perturbations remain consistent, except in 
cases where subdominant modes play a decisive role \cite{Zagli2024,Lucarini2025}.
These results pose a fundamental question regarding the interpretation of 
mathematical models of chaotic dynamics. Since multiple, structurally distinct 
models can yield virtually indistinguishable attractors and dynamical features, 
in what sense does any single model faithfully represent the underlying physical 
system? This observation suggests that the traditional objective of inferring 
governing equations may, in certain contexts, be of limited utility and potentially 
misleading. A more effective strategy may be to exploit data-driven approaches 
directly, for example through modern machine learning frameworks~\cite{KFGL:2021a,gauthier2021next,patel2021using,kim2023neural,flynn2023seeing,liu2024llms,zhang2024zero}.

%Intuitively, if two different evolution equations lead to closely corresponding chaotic attractor and Koopman representation, the two corresponding velocity fields on the attractor should agree with each other with high accuracy. We find that this is indeed the case for most of the points on the attractor, except for a small subsets where large deviations are observed as sudden spikes. Because of the local or impulsive nature of the deviations, they do not affect the values of long-term statistical quantities such as the Lyapunov exponents, nor the key features of the Koopman representation of the system. Note that agreement on the dominant Koopman modes implies that correlation between generic observables and response to generic perturbations should closely match, unless deep Koopman modes play an important role \cite{Zagli2024,Lucarini2025}. Our findings raise a fundamental question about the meaning of the mathematical models  of chaotic systems. Since many different models can produce a virtually indistinguishable attractor featuring almost equivalent dynamics, in what sense does an individual model reflect the physical reality? It is not unreasonable to speculate that finding the governing equations and making predictions about the dynamics of the underlying system may not be useful and could lead to misleading conclusions.  Instead, one should work directly with the available data taking advantage of, e.g., modern machine learning~\cite{KFGL:2021a,gauthier2021next,patel2021using,kim2023neural,flynn2023seeing,liu2024llms,zhang2024zero}.

\begin{figure} [ht!]
\centering
\includegraphics[width=0.8\linewidth]{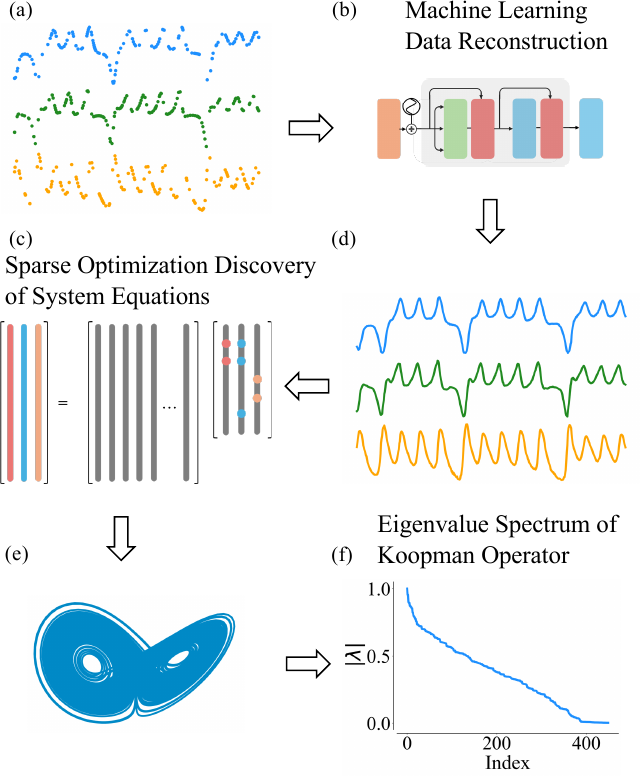}
\caption{Proposed method for assessing the accuracy and reliability of 
data-driven models of chaotic systems. 
(a) Representative time series with randomly missing observations.
(b) Machine-learning–based reconstruction of the complete time series.
(c–d) Sparse optimization applied to the reconstructed time series to identify 
governing equations.
(e) Reconstructed chaotic attractor.
(f) Eigenvalue spectrum of the Koopman operator estimated from the reconstructed series. The entire procedure is repeated for multiple random realizations of the missing data, often yielding system equations that differ substantially from one another.
%Proposed approach to assessing the accuracy and trustworthiness of the discovered system model in terms of its governing equations found from data. (a) Examples of time series with random missing observations. (b) a machine-learning scheme for reconstructing the complete time series. (c) Sparse optimization for finding the governing equations from the reconstructed time series in (d). (e) The reconstructed chaotic attractor. (f) The eigenvalues of the Koopman operator calculated from the reconstructed time series. The procedure is repeated for different random realizations of the missing observations, leading to the system equations that can be drastically different from each other.
}
\label{fig:Method}
\end{figure}

The proposed method to evaluate the accuracy and trustworthiness of the discovered 
system model in terms of its governing equations found from data is illustrated in 
Fig.~\ref{fig:Method}.  Suppose a target chaotic system produces continuous,
uniformly sampled time series. In experiments, measurement or observation 
may be disturbed, leading to random missing segments of the data, as shown in 
Fig.~\ref{fig:Method}(a). The missing data is imputed back into the time series 
by using, e.g., machine learning, as shown in Fig.~\ref{fig:Method}(b) [one
recent method~\cite{ZSL:2025} is described in Sec.~I of Supplementary 
Information (SI)]. This is the first step of a general 
procedure to optimally mitigate data gaps arising from measurement noise 
or device limitations. Sparse optimization~\cite{WYLKG:2011,BPK:2016} is then 
applied to the reconstructed time series to find the system equations, as 
illustrated in Figs.~\ref{fig:Method}(c) and \ref{fig:Method}(d), leading to a 
chaotic attractor [Fig.~\ref{fig:Method}(e)] that can be compared with the ground 
truth. Here, the basic idea of sparse optimization is that the evolution 
of the dynamical system can be represented using a minimal set of active terms, 
drawn from a predefined library with candidate basic functions (see Sec.~II of 
SI for more details). The system equations is further subject to the Koopman 
analysis, generating a spectrum of the Koopman eigenvalues, as illustrated in 
Fig.~\ref{fig:Method}(f).

Our Koopman analysis~\cite{Lucarini2025} is based on the numerical approximation 
of the Perron-Frobenius operator~\cite{LM:book}, which is the adjoint of the Koopman 
operator~\cite{klus2015numerical,ikeda2022koopman}. we utilize Ulam's 
method~\cite{froyland2007ulam, BFGM:2014} to discretize the state space and 
construct a finite-dimensional approximation of this operator, generate a long 
trajectory from the recovered equations after discarding an initial ``warm-up'' 
period to remove transients. Each dimension of the state space is normalized to 
the unit interval. The processed continuous state space is then partitioned into 
non-overlapping boxes using a uniform grid. Each point in the trajectory is 
assigned to its nearest box center and the number of one-step transitions between 
boxes is recorded to form a transition matrix $\mathbb{T}_m$, where the element 
$\mathbb{T}_m(i,j)$ counts the points of transition from box $i$ to $j$. Empty
boxes are removed and the matrix is row-normalized to form a stochastic transition 
matrix $\mathbb{M}$, which approximates the Perron-Frobenius operator $\mathcal{P}$ 
($\mathcal{P}=\mathbb{M}^\intercal$). The leading eigenvector of $\mathcal{P}$, 
corresponding to the eigenvalue $\lambda_0=1$, represents the invariant density 
of the system, which is the long-term probability distribution in the state space. 
The remaining eigenvectors, corresponding to $|\lambda_k|<1$, characterize slowly 
decaying modes of the density. (More details about the Koopman analysis can be 
found in Sec.~II of SI.)

\begin{figure} [ht!]
\centering
\includegraphics[width=0.8\linewidth]{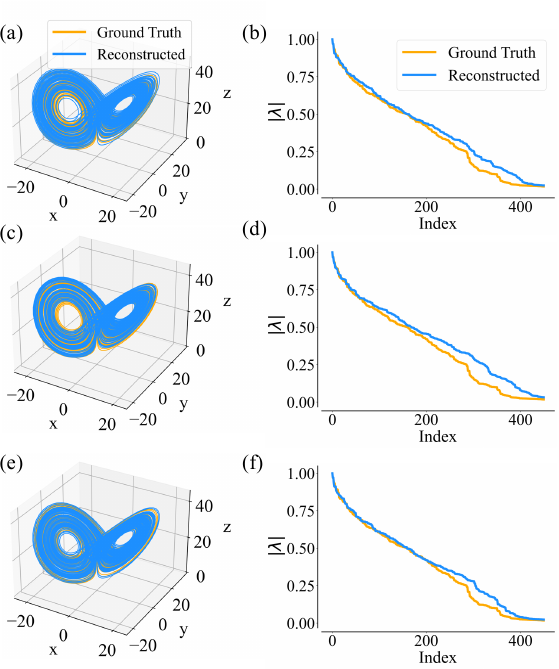}
\caption{Examples of the equations recovered from the chaotic Lorenz system and
the Koopman spectra. (a) For an arbitrary set of missing data at the missing ratio
of 20\%, the recovered system equations are (equation set \#1): 
$\dot{x} = -5.311x + 6.177y - 0.136xz + 0.113yz$, $\dot{y} = 25.167x - 0.927xz$, 
and $\dot{z} = 0.349y - 2.679z + 0.951xy$. The attractor generated by this set 
of equations is shown in blue, and the ground-truth Lorenz attractor is presented 
in orange. The two attractors agree well with each other: their KL distance is 
approximately 0.52. The Lyapunov exponents of the reconstructed attractor are 
approximately $(0.885, 0, -12.139)$, agreeing with the ground truth. (b) Koopman 
eigenvalue spectra of the recovered and original systems: the leading eigenvalues 
agree with each other with differences in smaller eigenvalues. (c,d) Same legends 
to (a,b) except that the missing data ratio is 30\%, where the recovered equations 
are (equation set \#2): $\dot{x} = -4.023x + 5.958y - 0.167xz + 0.113yz$, 
$\dot{y} = 28.361x - 3.051y - 1.026xz + 0.103yz$, and 
$\dot{z} = -0.105x + 0.176y - 2.691z + 0.942xy$. In this case, the KL distance 
is 0.42 and the Lyapunov exponents of the reconstructed attractor are 
approximately $(0.895, 0, -12.185)$. (e,f) Same legends to (a,b) except 
that the missing data ratio is 50\%. The recovered equations are 
(equation set \#3): $\dot{x} = -9.806x + 9.846y$, $\dot{y} = 24.176x - 0.901xz$, 
and $\dot{z} = -0.115x + 0.184y - 2.691z + 0.942xy$. The KL distance is 0.54 and 
the Lyapunov exponents of the reconstructed attractor are approximately 
$(0.907, 0, -13.404)$. In spite of the drastic differences among the three sets of 
equations reconstructed from different values of the missing data ratio, they 
produce essentially the same chaotic Lorenz attractor.}
\label{fig:Lorenz}
\end{figure}

A representative example is the classical chaotic Lorenz system given by: 
$\dot{x} = 10 (y-x)$, $\dot{y} = x (28 - z) - y$ and $\dot{z} = xy - (8/3)z$. 
For library determination, we assume strong prior knowledge 
and restrict the library to a minimal set of candidate functions: constant, 
first-order monomials, and their pairwise products. Consider the scenario of 
observing the system in a fixed time window but 20\% in total of the data points 
are missing at random times. Machine learning is then used to reconstruct the 
complete time series, from which the governing equations are found through 
sparse optimization. Figure~\ref{fig:Lorenz}(a) shows the chaotic attractor 
(blue) generated by the recovered equations (listed in the caption). For 
comparison, the ground-truth attractor from the original Lorenz equations is 
shown in orange. The two attractors agree well with each other, in spite of the 
vast difference between the reconstructed and the original Lorenz equations. 
Figure~\ref{fig:Lorenz}(b) presents the Koopman spectra calculated from the 
time series generated by the reconstructed equations and of the original 
Lorenz system. A large number of the dominant eigenvalues from the two spectra 
agree with each other, with differences beginning to emerge when the eigenvalues 
are below a small threshold. 

Similar features are observed for different scenarios of observations, as shown 
in Figs.~\ref{fig:Lorenz}(c,d) and Figs.~\ref{fig:Lorenz}(e,f) for missing data 
ratio 30\% and 50\%, respectively, where the physical observations may be deemed 
as ``poor.'' The striking phenomenon is that, regardless of how the different way 
of observing the system resulting in drastically different system equations, they 
produce the same ``correct'' chaotic attractor in terms of measures such as the 
Lyapunov exponents and the KL distance, as indicated in Fig.~\ref{fig:Lorenz}. 
Because missing data occurs randomly in time in the observational time window, in 
principle an infinite number of ``Lorenz'' systems can be found which generate 
essentially the same chaotic attractor in terms of measures including the Koopman 
eigenvalue spectrum. Even under the same sparsity setting but with 
different realizations of recovered data, the method produces a variety of distinct 
equations (See Sec.~IV A in SI). Six additional recovered equations for the Lorenz 
system, along with a statistical analysis, are presented in Sec.~III in SI. 
Moreover, two additional chaotic systems are tested with similar results, 
as reported in Sec.~III in SI. The Koopman modes of these systems are visualized 
in Sec.~IVB, demonstrating the close correspondence between the recovered equations 
and those of the original systems.

\begin{figure*} [ht!]
\centering
\includegraphics[width=0.9\linewidth]{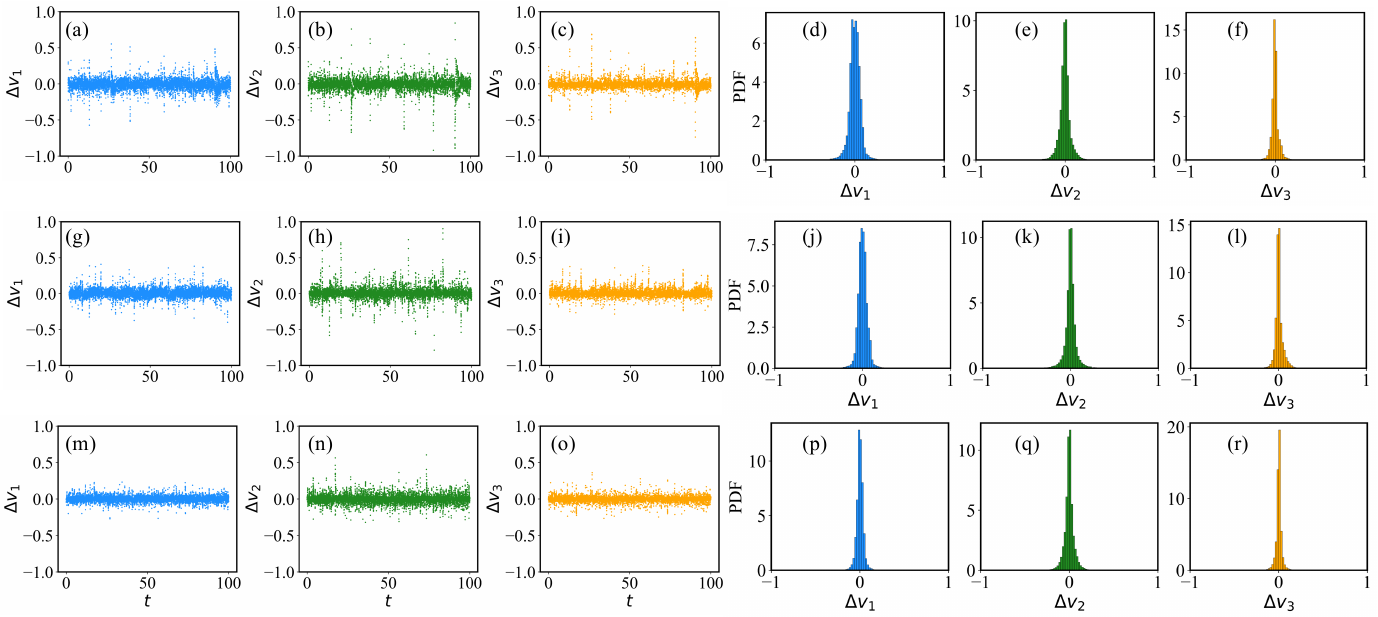}
\caption{On–off intermittency in velocity–field discrepancies. (a-c) The 
instantaneous component-wise discrepancies $\Delta v_i = (g_i-f_i)/\langle|f_i|\rangle$ 
between the recovered equations \#1 and ground truth Lorenz equations, exhibiting
long stretches of near-zero error (``off'' phases) and rare large deviations 
(``on'' bursts). (g-i) and (m-o) Same legends but for the recovered equation
sets \#2 and \#3, respectively. The corresponding probability density functions
of $\Delta \mathbf{v}_i$ are shown on the right side, which exhibit a sharp 
central peak at $\Delta \mathbf{v}_i\approx 0$ with diminishing values away from 
the center, confirming the on-off intermittency.}
\label{fig:On_Off}
\end{figure*}

How is it that drastically different equations can produce the same chaotic 
attractor? Intuitively, the velocity fields of the different sets of system
equations should agree with each other most of the time with rare deviations
so that the statistical properties of the resulting attractors are not affected.
An on-off intermittent behavior can then be anticipated. In particular, consider
two sets of equations. For the same point on the attractors, if the differences 
in the velocity fields are calculated as a function of time, most of the time 
the differences will be near zero. Indeed, the difference can be large but only 
occasionally. To demonstrate this, we choose the ground truth attractor as the 
reference with velocity field $\mathbf{f}$. Let $\mathbf{g}$ be the recovered 
velocity field. For point $k$ on the reference attractor with velocity 
$\mathbf{f}^k$, we search and identify $n$ points on the other attractor that 
are close to and compute their velocities $\{\mathbf{g}^j \}_{j=1}^n$. Due to 
the complex structure of the chaotic attractors, two close points could have 
different velocity fields. We thus select the point on those candidates that 
best aligns with $\mathbf{f}$ by maximizing the cosine similarity 
$\cos\theta_{jk}=\mathbf{f}^k\cdot \mathbf{g}^j/(\lVert \mathbf{f}^k\rVert\lVert \mathbf{g}^j\rVert)$, 
while also keeping the relative norm mismatch 
$|\lVert\mathbf{g}^k\rVert-\lVert\mathbf{f}^j\rVert|/\lVert\mathbf{f}^k\rVert$ 
as small as possible. Representative component-wise differences results are 
shown in Fig.~\ref{fig:On_Off}, which are characteristic of on-off intermittency. 
The difference of velocity fields $\Delta \mathbf{v}$ shown are normalized in 
each dimension, by $\mathbf{g}$ subtracting $\mathbf{f}$ and then dividing by 
the mean of $|\mathbf{f}|$. The time series of the velocity-field discrepancies 
reveal extended intervals where the component-wise velocity mismatch is near zero 
(``off'' state), interrupted by isolated bursts of large error (``on'' state). 
In the corresponding probability density functions, the sharp peak about zero 
and the tails from occasional large spikes explain how statistical invariants 
(e.g., Lyapunov exponents, KL distance, Koopman spectrum) remain unchanged despite 
algebraically distinct governing equations.

To summarize, we discovered a fundamental non-uniqueness in equation finding 
for chaotic dynamical systems. Reconstructing time series from sparse observations 
and applying sparse optimization to identify the governing equations will result 
in an infinite family of distinct equations, some even differing by extra terms 
or by sign. These equations nonetheless will generate statistically indistinguishable 
attractors, as confirmed by the Koopman analysis, KL distance, and Lyapunov 
exponents. In addition, on-off intermittency occurs in the differences in the 
velocity fields. The findings raise the question: if different equations can 
generate the same attractor, are these equations useful? Perhaps not. In fact,
working directly with the data without any system equations, e.g., by exploiting 
machine learning~\cite{Molinaetal:2023} and constructing digital 
twins~\cite{KWGHL:2023}, may be more effective. 

In general, machine learning can function as a universal approximator. For 
example, in regression tasks, richer parameterizations expand expressive power. 
While sparse regression may be effective for systems such as the Lorenz system 
and the additional examples (Sec.~IV in SI), their applicability diminishes for 
moderately more complex chaotic dynamics (e.g., a food-chain 
system~\cite{mccann1994nonlinear}), more intricate tasks (e.g., noise-induced 
transitions with SINDy~\cite{lin2024learning}), or real-world data. In such
cases, the ``true'' basis required to represent the dynamics is effectively 
unbounded. Consequently, sparse libraries become brittle, whereas high-capacity 
models such as Transformers and recurrent neural networks (e.g., reservoir 
computing and LSTMs) have demonstrated robust performance in 
forecasting~\cite{KFGL:2021a,gauthier2021next,patel2021using,kim2023neural,flynn2023seeing,liu2024llms,zhang2024zero}, denoising~\cite{zhai2023detecting} and 
controlling~\cite{canaday2021model,zhai2023model} nonlinear dynamical systems.

As we have demonstrated, it is possible to develop quantitative models by 
extracting equations from experimental data. The key criterion is that the 
resulting equations should be general, that is, independent of specific 
measurements. Indeed, such ``meaningful'' equations are highly valuable for 
mechanistic interpretability and principled extrapolation far beyond the 
training domain, as demonstrated in numerous 
studies~\cite{liu2021machine,tenachi2023deep,liu2024kan,PMBL:2025}. However, to 
ensure practical relevance, one must verify that the dynamical systems are 
governed by underlying physical laws and incorporate physics-informed 
techniques~\cite{udrescu2020ai} such as conservation laws, symmetries, and 
other structural priors, into the discovery process. 

This work was supported by AFOSR under Grant No.~FA9550-21-1-0438 and by ONR 
under Grant No.~N00014-24-1-2548. VL was supported by the Horizon Europe 
Projects ClimTIP (Grant No. 100018693) and Past2Future (Grant No. 101184070), 
and by ARIA SCOP-PR01-P003 - Advancing Tipping Point Early Warning AdvanTip.

%\bibliographystyle{ScienceAdvances}
%\bibliography{Modeling}

%apsrev4-2.bst 2019-01-14 (MD) hand-edited version of apsrev4-1.bst
%Control: key (0)
%Control: author (8) initials jnrlst
%Control: editor formatted (1) identically to author
%Control: production of article title (0) allowed
%Control: page (0) single
%Control: year (1) truncated
%Control: production of eprint (0) enabled
%
\end{document}